\documentclass[preprint,12pt]{elsarticle}

\usepackage{amssymb}
\usepackage{amsmath,bm}
\usepackage{graphicx}
\usepackage{hepunits}
\usepackage[svgnames]{xcolor}
\usepackage{multirow}
\usepackage{float}
\usepackage{textcomp}

\usepackage{hyperref}

\usepackage{shuffle}

\usepackage{cleveref}

\usepackage{url}

\usepackage{listings}
\definecolor{lightgray}{gray}{0.95}
\allowdisplaybreaks

\newcounter{bla}

\def\name{\texttt{FastGPL}}

\journal{Computer Physics Communications}

\begin{document}

\begin{frontmatter}



\title{\name{}: a C++ library for fast evaluation of generalized polylogarithms}


\author[a]{Yuxuan Wang}
\author[b]{Li Lin Yang\corref{c}}
\author[a]{Bin Zhou}

\address[a]{School of Physics and State Key Laboratory of Nuclear Physics and Technology,
Peking University, Beijing 100871, China}
\address[b]{Zhejiang Institute of Modern Physics, Department of Physics, Zhejiang University, Hangzhou 310027, China}

\cortext[c]{Corresponding author.\\\textit{E-mail address:} yanglilin@zju.edu.cn}

\begin{abstract}
We present \name{}, a C++ library for the fast evaluation of generalized polylogarithms which appear in many multi-loop Feynman integrals. We implement the iterative algorithm proposed by Vollinga and Weinzierl in a two-step approach, i.e., we generate concise expressions using an external program and hard-code them into the numeric library.
This allows efficient and accurate numeric evaluations of generalized polylogarithms suitable for Monte Carlo integration and event generation. Floating-point arithmetics are carefully taken care of to avoid loss of accuracy. As an application and demonstration, we calculate the two-loop corrections for Higgs boson production in the vector boson fusion channel at electron-positron colliders. \name{} is expected to be useful for event generators at the next-to-next-to-leading order accuracy.
\end{abstract}


\end{frontmatter}


\newpage
{\bf PROGRAM SUMMARY}

\begin{small}
\noindent
{\em Program Title: }\name{} \\
{\em Developer's repository link: }\url{https://github.com/llyang/FastGPL} \\
{\em Code Ocean capsule: }(to be added by Technical Editor) \\
{\em Licensing provisions (please choose one): }LGPL \\
{\em Programming language: }C++ \\
{\em Supplementary material:} \\
{\em Journal reference of previous version: } \\
{\em Does the new version supersede the previous version?: } \\
{\em Reasons for the new version: } \\
{\em Summary of revisions: } \\
{\em Nature of problem: }Fast numerical evaluation of generalized polylogarithms~\cite{Goncharov}.\\
{\em Solution method: }A two-step implementation of the algorithm of \cite{Vollinga}, where hard-coded expressions are generated by an external program.\\
{\em Additional comments including restrictions and unusual features: }Currently limited to GPLs up to weight 4.\\
   \\
{\raggedright
}
\end{small}

\newpage
\section{Introduction}

The generalized polylogarithms (GPLs, also called multiple polylogarithms, i.e., MPLs) \cite{Goncharov:1998kja} appear in many cutting-edge calculations of Feynman integrals and scattering amplitudes. In order to use these amplitudes to compute differential cross sections or to generate Monte Carlo event samples, it is highly desirable to obtain numeric values of GPLs as efficient as possible. A general algorithm for the numeric evaluation of GPLs has be proposed in \cite{Vollinga:2004sn}. An implementation of this algorithm has been built into the symbolic computation library \texttt{GiNaC} \cite{Bauer:2000cp, Vollinga:2005pk}. However, by default \texttt{GiNaC} is not suitable for fast numeric calculations since it is designed to work best with numbers of unlimited precision (as implemented in the \texttt{CLN} library).

A reimplementation of the algorithm using \texttt{Fortran} has appeared recently under the name \texttt{handyG} \cite{Naterop:2019xaf}. It is designed for the purpose of fast numeric evaluation suitable for Monte Carlo event generations. It has since then been widely used in multi-loop calculations as a fast alternative to \texttt{GiNaC}. However, during our usage of it, we found some drawbacks of the program which motivated us to write our own private implementation of the same algorithm. This gradually grows up into the library \name{} presented in this work.

The main drawbacks of \texttt{handyG} that we found can be summarized in the following: \emph{1)} The algorithm of \cite{Vollinga:2004sn} is an iterative one. If one naively follows the iterative approach, there can be situations where some functions are computed many times for the evaluation of one particular GPL. This greatly slows down the evaluation for such GPLs. For example, the average evaluation time for a single weight-4 GPL on a modern computer is about $10^{-4}$ seconds. However, as we will see in Section~\ref{sec:fastgpl}, the evaluation time of certain weight-4 GPLs can be as long as $\mathcal{O}(1)$ second. \emph{2)} The series representation of GPLs may sometimes lead to terms exceeding the maximal value that can be hold by a double-precision floating-point number. When this happens, \texttt{handyG} simply truncates the series, which leads to a loss of accuracy. This problem can of course be circumvented by using the quad-precision version of \texttt{handyG}. This however slows down the computation much more than necessary, if one only needs the final result in double-precision. \emph{3)} Partly due to the iterative nature of the algorithm, floating-point cancellations among different contributions often happen. In certain cases this leads to non-negligible losses of accuracy without careful treatment.

The aim of \name{} is avoid the above problems as much as possible in a floating-point implementation of the algorithm of \cite{Vollinga:2004sn}, while providing the overall efficiency and accuracy similar to \texttt{handyG}.
This paper is organized as follows. In Section~\ref{sec:notation}, we briefly introduce the notations and some important properties of GPLs. In Section~\ref{sec:fastgpl}, we describe the program \name{}, and provide some validations and benchmarks. In Section~\ref{sec:eenunuH} we discuss a working application in the two-loop corrections for the $e^+e^- \to \nu\bar{\nu}H$ scattering process. Finally we conclude in Section~\ref{sec:summary}. In the Appendix, we describe how to obtain, install and use \name{}.

\section{Notations and properties of GPLs}
\label{sec:notation}

In this section, we setup our notations and introduce some important properties of GPLs which will be used in their numeric computation. For a more comprehensive review, we refer the interested readers to Refs.~\cite{Duhr:2014woa}.
GPLs are complex-valued functions depending on a sequence of complex parameters $a_1,\ldots,a_n$ (we will often collectively refer to this sequence as $\bm{a}$) as well as a complex argument $z$. A GPL can be defined recursively as
\begin{align}
G(\bm{a}; z) = G(a_1,\ldots,a_n; z) &\equiv \int_{0}^{z} \frac{dt_1}{t_1-a_1} G(a_2,\ldots,a_n; t_1) \, .
\end{align}
with the initial kernels
\begin{align}
\quad G(; z) \equiv 1 \, , \quad G(\underbrace{0,\ldots,0}_{n}; z)
\equiv G(\bm{0}_n; z) \equiv \frac{\log^n z}{n!} \, ,
\end{align}
where $\bm{0}_n$ denotes a sequence of $n$ zeros. We will call the parameters $a_i$'s as the indices of the GPL. The number of indices is called the weight of the GPL. 

The GPLs satisfy a set of relations called the shuffle algebra. The product of a weight-$n$ GPL and a weight-$m$ GPL can be expressed as the sum of a number of weight-$(n+m)$ GPLs as following:
\begin{equation}
G(\bm{a}; z) \, G(\bm{b}; z) = \sum_{\bm{c} \in \bm{a} \shuffle \bm{b}} G(\bm{c}; z) \, ,
\end{equation}
where $\bm{a} = \{a_1,\ldots,a_n\}$, $\bm{b} = \{b_1,\ldots,b_m\}$, and $\bm{a} \shuffle \bm{b}$ denotes the shuffle product of the two sequences $\bm{a}$ and $\bm{b}$. Briefly speaking, an element in the shuffle product is a rearrangement of the entries in $\bm{a}$ and $\bm{b}$, such that the relative orders within $\bm{a}$ and $\bm{b}$ are preserved. As an example, considering the case $n=m=2$, we have
\begin{align}
\bm{a} \shuffle \bm{b} = \big\{ \{ a_1,a_2,b_1,b_2 \} \, , \{a_1,b_1,a_2,b_2\} \, , \{a_1,b_1,b_2,a_2\} \, , \nonumber
\\
\{b_1,a_1,a_2,b_2\} \, , \{b_1,a_1,b_2,a_2\} \, , \{b_1,b_2,a_1,a_2\} \big\} \, .
\end{align}

GPLs with the last index $a_n = 0$ are logarithmic divergent when $z \to 0$, which prevents a series representation in that region. The shuffle algebra can be used to remove these ``trailing zeros'' and to extract the divergences. For example, we have
\begin{equation}
G(a_1,a_2,0; z) = \log(z) \, G(a_1,a_2; z) - G(0,a_1,a_2; z) - G(a_1,0,a_2; z) \, .
\end{equation}
While this step can be taken care of numerically, it is strongly suggested to perform it analytically before feeding the expression into a numeric program. This avoids the unnecessary overhead in the numeric evaluation, and also allows one to transform the argument $z$ of the remaining GPLs to a non-negative real number (to be discussed later).

The evaluation of GPLs without trailing zeros boils down to computing the partial sum of their series representation. For that purpose, we first introduce the compressed notation of GPLs:
\begin{equation}
G_{m_1,\ldots,m_k}(a_1,\ldots,a_k; z) \equiv G(\bm{0}_{m_1-1},a_1,\ldots,\bm{0}_{m_k-1},a_k; z) \, ,
\end{equation}
where $a_1,\ldots,a_k$ are non-zero. Such a GPL can be expanded as
\begin{equation}
G_{m_1,\ldots,m_k}(a_1,\ldots,a_k; z)
= (-1)^k \sum_{i_1>\cdots>i_k>0} \frac{\left(\frac{z}{a_1}\right)^{i_1}}{i_1^{m_1}} \frac{\left(\frac{a_1}{a_2}\right)^{i_2}}{i_2^{m_2}} \cdots \frac{\left(\frac{a_{k-1}}{a_k}\right)^{i_k}}{i_k^{m_k}} \, ,
\label{eq:G_series}
\end{equation}
which converges under the conditions $|z| \leq |a_i|, (i=1,\ldots,k)$, and if $m_1 = 1$, one requires in addition $z \neq a_1$ (otherwise the GPL is logarithmic divergent). If these conditions are satisfied, we call the corresponding GPL as a ``convergent'' GPL. In a numeric implementation of Eq.~\eqref{eq:G_series}, one truncates the infinite series such that the remaining terms are negligible for a chosen accuracy. Already at this point, one can imagine that, for certain values of the indices and the argument, the powers of ratios in Eq.~\eqref{eq:G_series} can exceed the maximal value allowed in a given floating-point representation. Whether this may happen in practice is something we'll discuss in the next section.

It is easy to realize that not all GPLs are convergent in the sense of Eq.~\eqref{eq:G_series}. The main idea of the algorithm in \cite{Vollinga:2004sn} is to use the integral representation and the properties of GPLs to transform a given non-convergent GPL to a linear combination of convergence ones. We are not going to describe the algorithm in detail but refer the interested readers to \cite{Vollinga:2004sn, Naterop:2019xaf}. It suffices to mention that, one may encounter the same (convergent or non-convergent) GPL many times in this multi-step transformation, which then needs to be evaluated many times if the algorithm is implemented recursively.

A GPL without trailing zeros satisfies another useful relation:
\begin{equation}
G(a_1,\ldots,a_m; z) = G(\lambda a_1,\ldots,\lambda a_m; \lambda z) \, ,
\end{equation}
for any non-zero complex number $\lambda$. This property can be used to convert the argument $z$ of GPLs to be a non-negative real number. This allows the program to easily decide which side of a branch cut to take according to the signs of the imaginary parts of the indices. This is important since quite often the imaginary parts of the parameters are infinitesimal ones coming from the Feynman $+i\epsilon$ prescription. Setting a small imaginary part instead of an infinitesimal one is sometimes unacceptable in a numeric computation. Therefore, in \name{} we require the input argument $z$ to be a non-negative real numbers, and allow the users to indicate the signs of the infinitesimal imaginary parts with an extra set of input parameters. This convention is the same as \texttt{GiNaC}.

\section{The program library \name{}}
\label{sec:fastgpl}

\subsection{Implementation}

The implementation of \name{} is a two-step process. An in-house program in \texttt{Mathematica} is built up to perform the transformation from non-convergent GPLs to convergent ones. Note that the transformation is different for different relative sizes of the indices $\{a_i\}$, and the program needs to generate an expression for each individual case. The resulting expressions consisting of convergent GPLs and a few lower-weight non-convergent ones are then converted to \texttt{C++}. Abbreviations are introduced at this step such that each function in the expression for a given GPL is computed only once. The \texttt{Mathematica} code is not fully automatic at the moment, and therefore only GPLs up to weight-4 have been currently implemented.

Avoiding the repeated evaluations of the same GPLs can lead to significant performance gain in certain cases. 
Take
$$\mathtt{G(1.0025, 0.989, 0.45, 0.89+0.24i; 1)}$$
as an example. In an iterative implementation of the algorithm, various weight-3 and weight-4 GPLs need to be recursively evaluated for $\mathcal{O}(500000)$ times. This significantly slows down the computation, and it takes \texttt{handyG} about 2 seconds to evaluate the above GPL\footnote{There is an undocumented compilation option (disabled by default) of \texttt{handyG} which enables a cache system for convergent GPLs. This may slightly reduce the evaluation time sometimes but does not work consistently. We therefore leave that option off in our tests.}. On the other hand, the evaluation time of \name{} is less than 0.01 seconds. The non-iterative implementation of \name{} also reduces the accuracy losses due to floating-point arithmetics, and the resulting precision is usually better or at least comparative. Here and in the following, all benchmarks are performed using a single core of an Intel Xeon E5-1680v3 CPU with the gcc 9.4 compiler under Ubuntu Linux 18.04.

After transforming to convergent GPLs, they can then be computed using the series representation \eqref{eq:G_series}.  While this is a trivial exercise for a program using unlimited precision numbers (such as \texttt{GiNaC}), it is a bit challenging for a program using floating point numbers. The problem is that although the terms in Eq.~\eqref{eq:G_series} gradually become smaller order-by-order, some of the powers $(a_{j-1}/a_j)^{i_j}$ may become very large at high orders. They could exceed the maximal value that can be stored in a floating-point format (which is about $10^{308}$ for double-precision). One may truncate the series before that happens, which however leads to a loss of accuracy. One may also use a floating-point format with more bits which can represent numbers in a bigger range. Such an option is provided in \texttt{handyG}, where 128-bit quad-precision numbers can be used throughout. This however significantly slows down the computation, which is certainly not desirable if one only needs the final results in double-precision. \name{} approaches this problem slightly differently: it switches to a higher-precision format (long double at the moment, which however is compiler-dependent) only when necessary, i.e., when it detects the possibility of overflow.

\begin{table}[t!]
\newcommand{\tabincell}[2]{\begin{tabular}{@{}#1@{}}#2\end{tabular}}
\centering
\begin{tabular}{|c|c|c|}
\hline
& $g_1 \times 10^{5}$ & $g_2 \times 10^{5}$
\\ \hline
\texttt{GiNaC} & {\small \tabincell{c}{$\mathtt{3.6870217689293}$ \\$\mathtt{- 7.5295211195637 i}$}} & {\small \tabincell{c}{$\mathtt{1.01839797324439}$ \\$\mathtt{- 1.50327528421384 i}$}}
\\ \hline
\name{} & {\small \tabincell{c}{$\mathtt{3.6870217689292}$ \\$\mathtt{- 7.5295211195637 i}$}} & {\small \tabincell{c}{$\mathtt{1.01839797324439}$ \\$\mathtt{- 1.50327528421384 i}$}}
\\ \hline
\texttt{handyG} & {\small \tabincell{c}{$\mathtt{3.6874023785452}$ \\$\mathtt{- 7.5284903971195 i}$}} & {\small \tabincell{c}{$\mathtt{1.01819546357542}$ \\$\mathtt{- 1.50329421576427 i}$}}
\\ \hline
\end{tabular}
\caption{A demonstration of accuracy-loss due to floating-point overflow occurred in \texttt{handyG} but not in \name{}.\label{tab:overflow}}
\end{table}

The above approach has been tested at millions of randomly sampled points, and also in realistic applications. No truncation of the series is observed. As a demonstration, we show in Table~\ref{tab:overflow} the results for the following two GPLs:
\begin{align*}
g_1 &= \mathtt{G(-223.7-576.0i, -0.29-0.74i, -0.29-0.74i, 0.26+0.67i; 1)} \, ,
\\
g_2 &= \mathtt{G(0.24+0.61i, -0.21-0.55i, 0, -4735.0-12190.9i; 1)} \, .
\end{align*}
It can be seen that the results of \name{} agree perfectly with those of \texttt{GiNaC}. On the other hand, the results of \texttt{handyG} deviate relatively by as much as $10^{-4}$. Note that the above GPLs appear in the calculation of a two-loop scattering amplitude to be presented later. Hence it is not just an academic study but have practical consequences.

\subsection{Benchmarks}

We have performed a thorough test of \name{} at weights 1, 2, 3 and 4 with randomly generated parameters. The relative precision of weight-4 GPLs is typically at the level of $10^{-13}$ or better. The lower weight GPLs are of course more accurate. We believe that this is enough for most phenomenological applications.

\name{} is built with efficiency in mind. In the following we show some benchmarks of it in the computations of weight-4 GPLs. The number of iterations needed for a GPL depends crucially on the absolute values of the non-zero indices $\{a_i\}$ compared to the argument $x$. Hence we classify weight-4 GPLs into several categories accordingly, and investigate the typical time required for their evaluations. We will use lowercase letters to label ``small'' indices, i.e., $|a| < |x|$; and use uppercase letters to label ``big'' indices, i.e., $|A| \geq |x|$. For example, $0abC$ denotes a category of weight-4 GPLs with one zero-index, two small indices and a big index. Note that the order of the indices is irrelevant in this classification.

\begin{table}[t!]
	\centering
	\begin{tabular}{|l|c|c|c|}
		\hline
		& $t_{\text{f}}$~(s) & $t_{\text{h}}$~(s) & $t_{\text{h}}/t_{\text{f}}$ \\
		\hline
		$\mathtt{G(1.0025, 0.989, 0.45, 0.89+0.24i; 1)}$  & $0.006$ & $2.2$ & $\sim 400$ 
		\\ \hline
		$\mathtt{G(0.998, 1.0545+0.127i, 0.91+0.25i, -0.226; 1)}$  & $ 0.004 $ & $ 1.5 $ & $\sim 400$ 
		\\ \hline
		$\mathtt{G(-1.04, -0.97, 0.25, -0.84+0.45i; 1)}$  & $0.004$ & $1.1$ & $\sim 300$
		\\ \hline
	\end{tabular}
	\caption{Average evaluation times of several GPLs which require many iterations.\label{tab:slow}}
\end{table}

\begin{table}[t!]
	\centering
	\begin{tabular}{|c|c|c|c|c|c|}
		\hline
		& $0aBC$ & $0abC$ & $0abc$ & $00aB$ & $00ab$
		\\ \hline
		$t_{\text{f}}$~(ms)  & $0.22$ & $0.25$ & $0.20$ & $0.08$ & $0.05$ 
		\\ \hline
		$t_{\text{h}}$~(ms)  & $3.1$ & $5.8$ & $4.5$ & $1.3$ & $0.80$
		\\ \hline
		$t_{\text{h}}/t_{\text{f}}$  & $\sim 14$ & $\sim 23$ & $\sim 23$ & $\sim 17$ & $\sim 16$
		\\ \hline
	\end{tabular}
	\begin{tabular}{|c|c|c|c|c|}
		\hline
		& $ABCD$ & $abCD$ & $abcD$ & $abcd$ 
		\\ \hline
		$t_{\text{f}}$~(ms) & $0.22$ & $0.47$ & $0.50$ & $0.42$ 
		\\ \hline
		$t_{\text{h}}$~(ms) & $1.7$ & $7.4$ & $11.0$ & $9.1$ 
		\\ \hline
		$t_{\text{h}}/t_{\text{f}}$ & $\sim 7.5$ & $\sim 16$ & $\sim 22$ & $\sim 22$
		\\ \hline
	\end{tabular}
	\caption{Average evaluation times of a few categories of weight-4 GPLs.\label{tab:more}}
\end{table}

In Table~\ref{tab:slow}, we first show several examples of GPLs that require many iterations in the algorithm. The average evaluation times using \name{} and \texttt{handyG} are denoted as $t_f$ and $t_h$, respectively. Apparently these are cases where \name{} mostly benefits from the preprocessing of the iterative algorithm, and the performance gain is quite impressive. This of course does not represent the generic behaviors of the two programs. In Table~\ref{tab:more} we show the average evaluation times of a few categories of weight-4 GPLs, where the indices are randomly sampled within each category. One can see that \name{} is typically faster by $\sim 20$ times in these situations. In the next section, we apply our program to a realistic calculation of a Higgs production process, to demonstrate its efficiency in practice.

\section{Application to $e^+e^- \to \nu_e \bar{\nu}_e W^* W^* \to \nu_e \bar{\nu}_e H$}
\label{sec:eenunuH}

\begin{figure}[t!]
	\centering
	\includegraphics[width=0.5\textwidth]{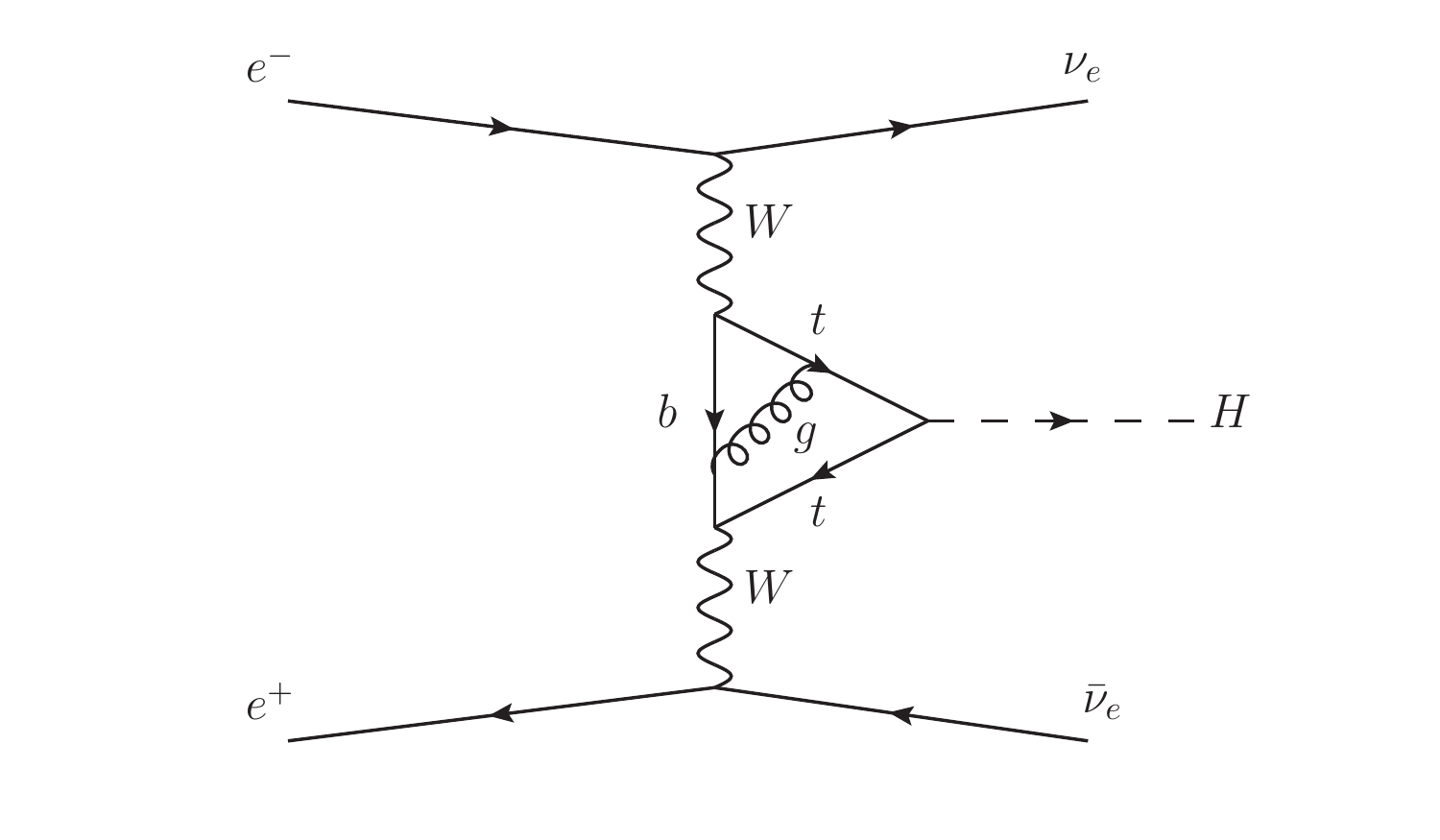}
	\vspace{-3ex}
	\caption{\label{figure:diagram}A typical Feynman diagram contributing to the mixed QCD-EW corrections for the Higgs production process via $W$ boson fusion.}
\end{figure}

Vector boson fusion is one of the dominant Higgs production channels at a future electron-positron collider as a Higgs factory. The expected high experimental precision demands improvements on theoretical calculations for this process. The leading order (LO) and next-to-leading order (NLO) electroweak (EW) corrections for the total and differential cross sections are available in \cite{Altarelli:1987ue, Kilian:1995tr,Belanger:2002me,Denner:2003yg, Denner:2003iy,Denner:2004jy}. Here, we add to those results the next-to-next-to-leading order (NNLO) mixed QCD-EW corrections at order $\alpha\alpha_s$. A typical Feynman diagram at this order is shown in Fig.~\ref{figure:diagram}.

In our calculation we neglect masses of all light fermions except that of the top quark. We generate the two-loop amplitude using \texttt{FeynArts} \cite{Hahn:2000kx}, and manipulate the resulting expressions using \texttt{Mathematica}. The scalar integrals are reduced to a set of master integrals using \texttt{FIRE6} \cite{Smirnov:2019qkx} and \texttt{LiteRed} \cite{Lee:2013mka}. The expressions for all the master integrals in terms of GPLs can be found in \cite{DiVita:2017xlr, Ma:2021cxg}. We renormalize the fields and the masses in the on-shell scheme, while the fine structure constant $\alpha$ is defined in the $G_\mu$ scheme:
\begin{equation}
\alpha_{G_{\mu}} = \frac{\sqrt{2}}{\pi} G_{\mu} m_W^2 \left( 1 - \frac{m_W^2}{m_Z^2} \right) .
\end{equation}
The standard model parameters are chosen as \cite{ParticleDataGroup:2020ssz}: $m_t=\unit{172.76}{\GeV}$, $m_H=\unit{125.1}{\GeV}$, $m_Z=\unit{91.1876}{\GeV}$, $m_W=\unit{80.379}{\GeV}$, $\alpha_s(m_Z)=0.1179$ and $G_{\mu}=\unit{$1.1663787 \times 10^{-5}$}{\GeV^{-2}}$. The default renormalization scale for $\alpha_s$ is chosen as the center-of-mass energy $\mu=\sqrt{s}$.

\begin{table}[t!]
	\centering
	\begin{tabular}{|c|c|c|c|c|c|c|}
		\hline
		$\sqrt{s}$~(GeV) & $\sigma_\text{LO}$~(fb) & $\delta\sigma_\text{NNLO}$~(fb) & $t_{\text{f}}$~(h) & $t_{\text{h}}$~(h) & $ t_{\text{h}}/t_{\text{f}} $
		\\ \hline
		$250$  & $7.88$ & $0.010$ & $0.45$ & $8.60$ & $\sim 19$
		\\ \hline
		$350$  & $30.6$ & $0.040$ & $0.51$ & $9.02$ & $\sim 18$
		\\ \hline
		$500$  & $74.8$ & $0.101$ & $0.52$ & $9.24$ & $\sim 18$
		\\ \hline
	\end{tabular}
	\caption{Results for the total cross sections at three center-of-mass energies evaluated using \name{} and \texttt{handyG}.}\label{tab:total}
\end{table}

With the above knowledge, we can readily compute the total and differential cross sections by integrating over the phase-space. The two-loop amplitude contains about 8000 GPLs, most of which are of weight 4. In a Monte Carlo integrator, these GPLs need to be evaluated at randomly generated phase-space points. For the total cross section, we sample 10,000 points using the VEGAS algorithm implemented in the library \texttt{Cuba} \cite{Hahn:2004fe,Hahn:2016ktb} for each of the center-of-mass energies $\sqrt{s}=\unit{250}{\GeV}$, $\unit{350}{\GeV}$ and $\unit{500}{\GeV}$. The Monte Carlo uncertainties are at the level of a few percents, and for better accuracies one should increase the number of sample points. The computation is done using both \name{} and \texttt{handyG}, and the results agree with each other within the integration uncertainties. We also compute the LO cross sections using \texttt{FeynCalc} \cite{Shtabovenko:2016olh} in \texttt{Mathematica}. The results are shown in Table~\ref{tab:total}. It is perhaps unsurprising that \name{} is approximately 20 times faster, which demonstrates its power in practical applications.

\begin{figure}[t!]
	\centering
	\includegraphics[width=0.48\textwidth]{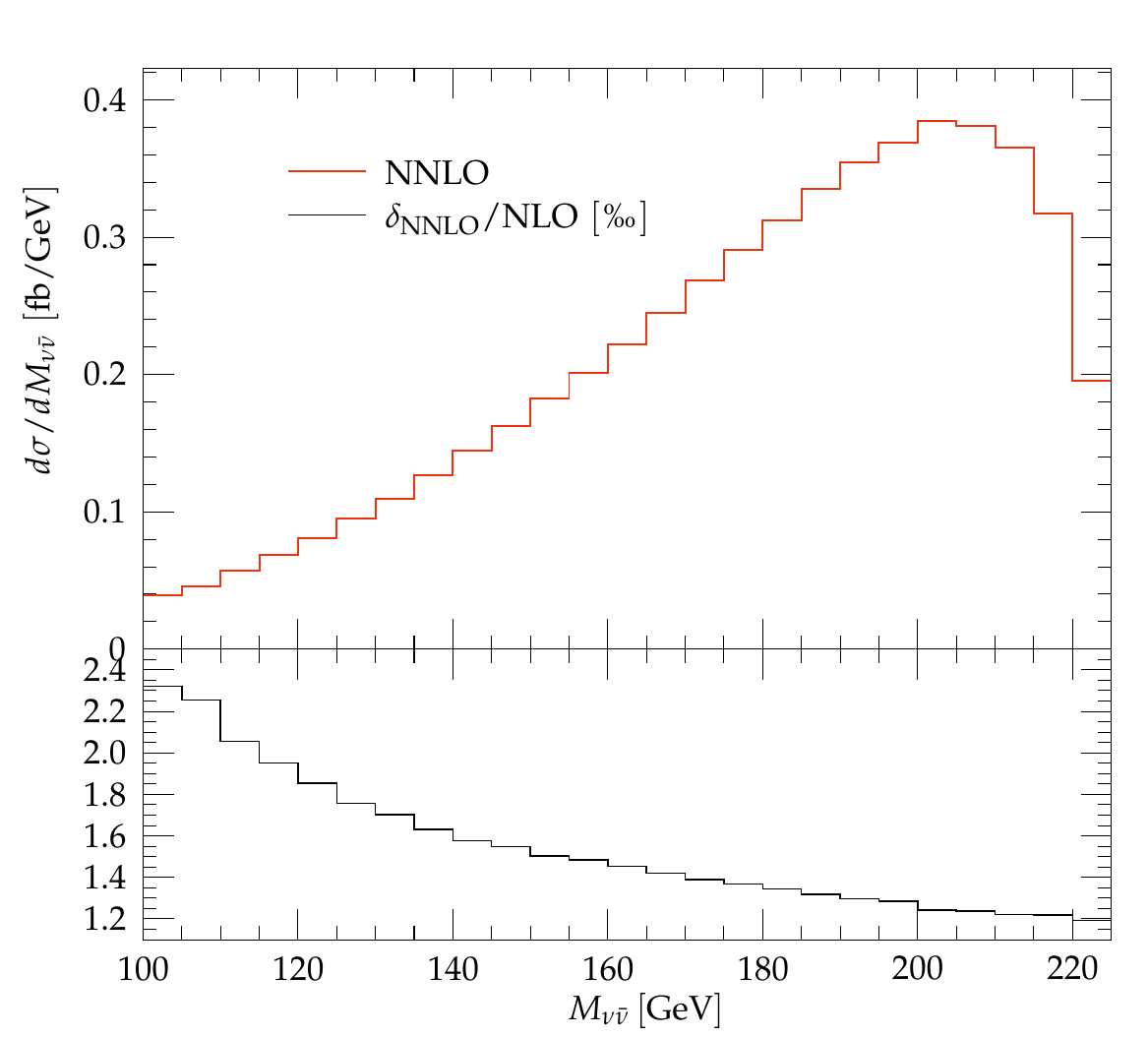}
	\includegraphics[width=0.48\textwidth]{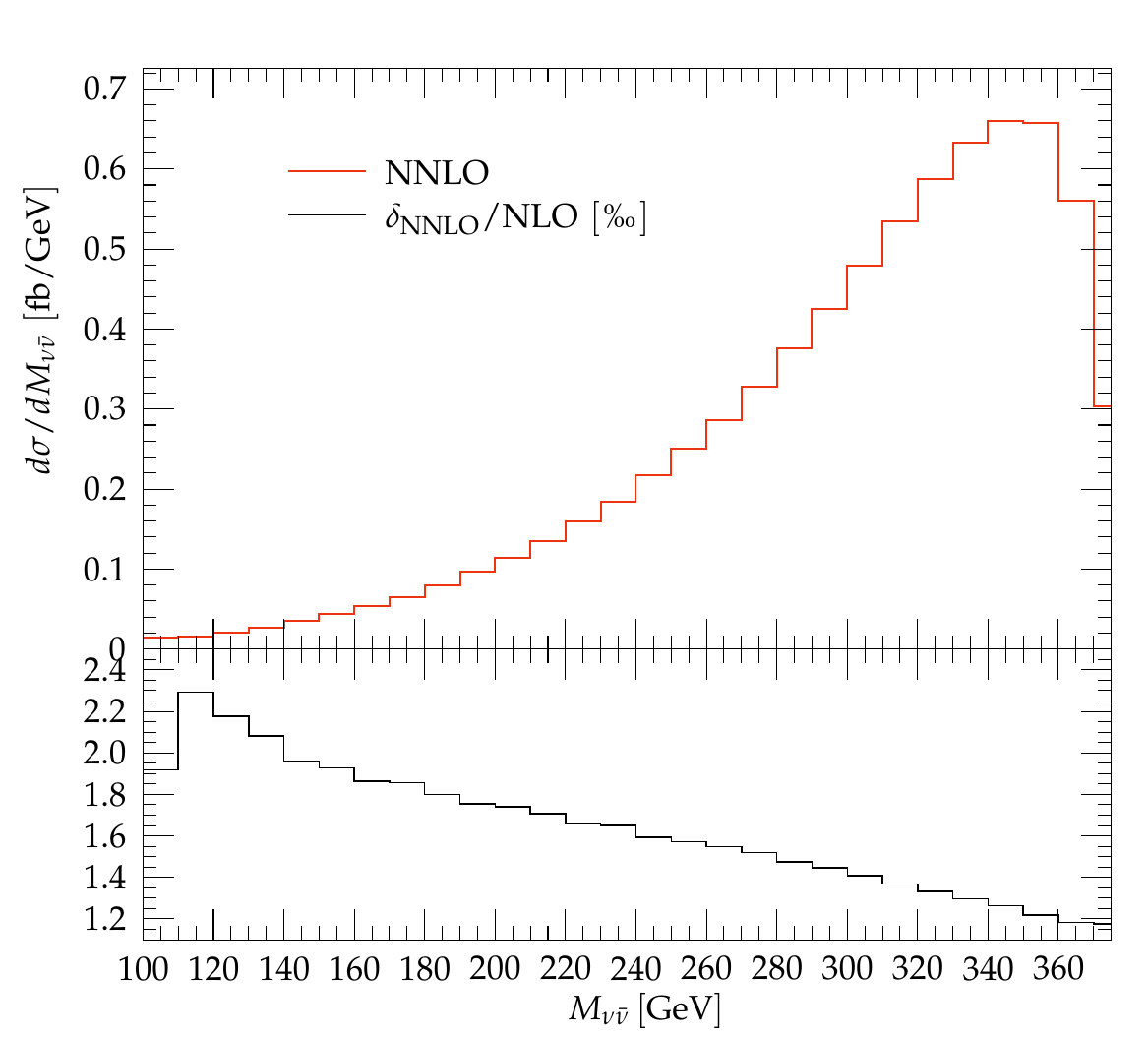}
	\vspace{-1ex}
	\caption{\label{figure:Gmus34}The $M_{\nu\bar{\nu}}$ distribution at $\sqrt{s}=\unit{350}{\GeV}$ (left) and $\unit{500}{\GeV}$ (right).}
\end{figure}

We now proceed to study the differential cross sections. Here we need much more sample points for acceptable integration uncertainties. Hence we will use \name{} exclusively. The $\nu_e \bar{\nu}_e H$ final state can actually come from two subprocesses: $ZH$ associated production with the $Z$ boson decaying to neutrinos, and the vector boson fusion channel studied in this section. An efficient way to disentangle the two subprocesses is to measure the invariant mass $M_{\nu\bar{\nu}}$ of the two neutrinos, which can be done by measuring the momenta of all visible particles and computing the missing energy. In the $ZH$ subprocess $M_{\nu\bar{\nu}}$ is located near the $Z$ boson mass, while in the vector boson fusion subprocess $M_{\nu\bar{\nu}}$ tends to be larger when there's enough collider energy. In Fig.~\ref{figure:diagram} we depict the $M_{\nu\bar{\nu}}$ distribution for $\sqrt{s}=\unit{350}{\GeV}$ and $\unit{500}{\GeV}$ with a lower cut at $\unit{100}{\GeV}$. For the NLO differential cross sections we employ the automated generator \texttt{Sherpa}~\cite{Kallweit:2014xda, Sherpa:2019gpd} with \texttt{OpenLoops}~\cite{Buccioni:2019sur} for the computation of one-loop integrals. The NNLO corrections are at the level of a few per-mille, which could be be measurable at a future high precision Higgs factory.

\section{Summary and Outlook}
\label{sec:summary}

In this paper, we present \name{}, a C++ library for the fast evaluation of GPLs which appear in many multi-loop Feynman integrals. We implement the iterative algorithm proposed by Vollinga and Weinzierl in a two-step approach, i.e., we generate concise expressions using an external program and hard-code them into the numeric library.
This allows efficient and accurate numeric evaluations of generalized polylogarithms suitable for Monte Carlo integration and event generation. Floating-point arithmetics are carefully taken care of to avoid loss of accuracy. At the moment, the library only supports GPLs up to weight 4, and higher weight GPLs will be added in the future.

As an application and demonstration, we calculate the NNLO mixed QCD-EW corrections for Higgs boson production in the vector boson fusion channel at electron-positron colliders. We find that \name{} is about 20 times faster on average than \texttt{handyG} in this application. Hence, \name{} is expected to be rather useful for event generators at the NNLO accuracy.

Analytic representations of more complicated Feynman integrals are expected to require functions beyond GPLs, such as (integrals over) elliptic integrals. For that \name{} provides fast numeric evaluations of the complete elliptic integrals of the first and second kinds. Their usage can be found in the Appendix. In the future we may also extend the library to calculate the elliptic multiple polylogarithms \cite{Adams:2016xah, Remiddi:2017har, Broedel:2017kkb, Broedel:2018qkq}.

\vspace{3ex}

\textbf{Acknowledgement.}
The authors would like to thank Yannick Ulrich and Stefan Weinzierl for the helps on \texttt{handyG} and \texttt{GiNaC}.
This work was supported in part by the National Natural Science Foundation of China under Grant No. 11975030 and 11635001.

\appendix

\section{Installation and usage}
\label{Appendix:install}

\name{} can be obtained from GitHub using
\begin{lstlisting}
git clone https://github.com/llyang/FastGPL.git
\end{lstlisting}
To compile and install the library, one needs to have CMake installed, and the following commands have to be executed:
\begin{lstlisting}
cd FastGPL
mkdir build && cd build
cmake -DCMAKE_INSTALL_PREFIX=/usr/local ..
make
make install
\end{lstlisting}
where the prefix can be chosen by the user. The compilation procedure also builds a simple program in the \texttt{test} directory, which demonstrates the usage of the library.

To use \name{}, one simply includes the header file \texttt{FastGPL.h}, and invokes the function \texttt{FastGPL::G} in the program. The syntax can be seen from the following code snippet:
\begin{lstlisting}
    vector<complex<double>> a { {0.1, 0.2}, 0.4 };
    vector<int> s {1, -1};
    double x { 0.8 };
    cout << FastGPL::G(a, s, x) << '\n';
    cout << FastGPL::G(a, x) << '\n';
\end{lstlisting}
Here, \texttt{x} must be a non-negative real number. The reason is given in Section~\ref{sec:notation}. In the \texttt{G(a, s, x)} version of the function, \texttt{a} is a vector of complex indices, and \texttt{s} is a vector with elements being either $\mathtt{+1}$ or $\mathtt{-1}$, which denote the signs of the imaginary parts of the corresponding indices. If one uses the $\texttt{G(a, x)}$ version of the function, and if the imaginary part of a certain index is zero, the program will automatically choose $\mathtt{+1}$ as the sign. This convention is the same as $\texttt{GiNaC}$. Hence, the last two lines in the above code snippet correspond to $\mathtt{G(0.1+0.2i, 0.4-i0; 0.8)}$ and $\mathtt{G(0.1+0.2i, 0.4+i0; 0.8)}$, respectively.

As a side effect, \name{} also provides several functions for the evaluation of normal polylogarithms $\mathrm{Li}_n(z)$ up to weight 8 and the Neilsen's generalized polylogarithm $\mathrm{S}_{2,2}(z)$ for the complex argument $z$. Furthermore, as elliptic integrals often appear in loop integrals which cannot be represented purely with GPLs, we also provide two functions to evaluate the complete elliptic integrals of the first and the second kinds. The usage of these functions can be seen from the following code snippet:
\begin{lstlisting}    
    complex<double> z { 1.2, 0.5 };
    int s { 1 };
    cout << FastGPL::PolyLog(6, z, s);
    cout << FastGPL::S22(z, s);
    cout << FastGPL::EllipticK(z);
    cout << FastGPL::EllipticE(z);
\end{lstlisting}
Note that for \texttt{PolyLog} and \texttt{S22}, \texttt{s} will default to $\mathtt{-1}$ if it's omitted and if the imaginary part of $\mathtt{z}$ is zero. This convention is the same as \texttt{GiNaC} and \texttt{Mathematica}.





\bibliographystyle{elsarticle-num}
\bibliography{cite}

\begin{thebibliography}{10}
\expandafter\ifx\csname url\endcsname\relax
  \def\url#1{\texttt{#1}}\fi
\expandafter\ifx\csname urlprefix\endcsname\relax\def\urlprefix{URL }\fi
\expandafter\ifx\csname href\endcsname\relax
  \def\href#1#2{#2} \def\path#1{#1}\fi

\bibitem{Goncharov:1998kja}
A.~B. Goncharov, {Multiple polylogarithms, cyclotomy and modular complexes},
  Math. Res. Lett. 5 (1998) 497--516.
\newblock \href {http://arxiv.org/abs/1105.2076} {\path{arXiv:1105.2076}},
  \href {http://dx.doi.org/10.4310/MRL.1998.v5.n4.a7}
  {\path{doi:10.4310/MRL.1998.v5.n4.a7}}.

\bibitem{Vollinga:2004sn}
J.~Vollinga, S.~Weinzierl, {Numerical evaluation of multiple polylogarithms},
  Comput. Phys. Commun. 167 (2005) 177.
\newblock \href {http://arxiv.org/abs/hep-ph/0410259}
  {\path{arXiv:hep-ph/0410259}}, \href
  {http://dx.doi.org/10.1016/j.cpc.2004.12.009}
  {\path{doi:10.1016/j.cpc.2004.12.009}}.

\bibitem{Bauer:2000cp}
C.~W. Bauer, A.~Frink, R.~Kreckel, {Introduction to the GiNaC framework for
  symbolic computation within the C++ programming language}, J. Symb. Comput.
  33 (2002) 1--12.
\newblock \href {http://arxiv.org/abs/cs/0004015} {\path{arXiv:cs/0004015}},
  \href {http://dx.doi.org/10.1006/jsco.2001.0494}
  {\path{doi:10.1006/jsco.2001.0494}}.

\bibitem{Vollinga:2005pk}
J.~Vollinga, {GiNaC: Symbolic computation with C++}, Nucl. Instrum. Meth. A 559
  (2006) 282--284.
\newblock \href {http://arxiv.org/abs/hep-ph/0510057}
  {\path{arXiv:hep-ph/0510057}}, \href
  {http://dx.doi.org/10.1016/j.nima.2005.11.155}
  {\path{doi:10.1016/j.nima.2005.11.155}}.

\bibitem{Naterop:2019xaf}
L.~Naterop, A.~Signer, Y.~Ulrich, {handyG \textemdash{}Rapid numerical
  evaluation of generalised polylogarithms in Fortran}, Comput. Phys. Commun.
  253 (2020) 107165.
\newblock \href {http://arxiv.org/abs/1909.01656} {\path{arXiv:1909.01656}},
  \href {http://dx.doi.org/10.1016/j.cpc.2020.107165}
  {\path{doi:10.1016/j.cpc.2020.107165}}.

\bibitem{Duhr:2014woa}
C.~Duhr, {Mathematical aspects of scattering amplitudes}, in: {Theoretical
  Advanced Study Institute in Elementary Particle Physics}: {Journeys Through
  the Precision Frontier: Amplitudes for Colliders}, 2015, pp. 419--476.
\newblock \href {http://arxiv.org/abs/1411.7538} {\path{arXiv:1411.7538}},
  \href {http://dx.doi.org/10.1142/9789814678766_0010}
  {\path{doi:10.1142/9789814678766_0010}}.

\bibitem{Altarelli:1987ue}
G.~Altarelli, B.~Mele, F.~Pitolli, {Heavy Higgs Production at Future
  Colliders}, Nucl. Phys. B 287 (1987) 205--224.
\newblock \href {http://dx.doi.org/10.1016/0550-3213(87)90103-9}
  {\path{doi:10.1016/0550-3213(87)90103-9}}.

\bibitem{Kilian:1995tr}
W.~Kilian, M.~Kramer, P.~M. Zerwas, {Higgsstrahlung and W W fusion in e+ e-
  collisions}, Phys. Lett. B 373 (1996) 135--140.
\newblock \href {http://arxiv.org/abs/hep-ph/9512355}
  {\path{arXiv:hep-ph/9512355}}, \href
  {http://dx.doi.org/10.1016/0370-2693(96)00100-1}
  {\path{doi:10.1016/0370-2693(96)00100-1}}.

\bibitem{Belanger:2002me}
G.~Belanger, F.~Boudjema, J.~Fujimoto, T.~Ishikawa, T.~Kaneko, K.~Kato,
  Y.~Shimizu, {Full O(alpha) corrections to e+ e- ---\ensuremath{>} nu anti-nu
  h by GRACE}, Nucl. Phys. B Proc. Suppl. 116 (2003) 353--357.
\newblock \href {http://arxiv.org/abs/hep-ph/0211268}
  {\path{arXiv:hep-ph/0211268}}, \href
  {http://dx.doi.org/10.1016/S0920-5632(03)80198-6}
  {\path{doi:10.1016/S0920-5632(03)80198-6}}.

\bibitem{Denner:2003yg}
A.~Denner, S.~Dittmaier, M.~Roth, M.~M. Weber, {Electroweak radiative
  corrections to single Higgs boson production in e+ e- annihilation}, Phys.
  Lett. B 560 (2003) 196--203.
\newblock \href {http://arxiv.org/abs/hep-ph/0301189}
  {\path{arXiv:hep-ph/0301189}}, \href
  {http://dx.doi.org/10.1016/S0370-2693(03)00370-8}
  {\path{doi:10.1016/S0370-2693(03)00370-8}}.

\bibitem{Denner:2003iy}
A.~Denner, S.~Dittmaier, M.~Roth, M.~M. Weber, {Electroweak radiative
  corrections to e+ e- ---\ensuremath{>} nu anti-nu H}, Nucl. Phys. B 660
  (2003) 289--321.
\newblock \href {http://arxiv.org/abs/hep-ph/0302198}
  {\path{arXiv:hep-ph/0302198}}, \href
  {http://dx.doi.org/10.1016/S0550-3213(03)00269-4}
  {\path{doi:10.1016/S0550-3213(03)00269-4}}.

\bibitem{Denner:2004jy}
A.~Denner, S.~Dittmaier, M.~Roth, M.~M. Weber, {Electroweak corrections to e+
  e- ---\ensuremath{>} f anti-f H}, Nucl. Phys. B Proc. Suppl. 135 (2004)
  88--91.
\newblock \href {http://arxiv.org/abs/hep-ph/0406335}
  {\path{arXiv:hep-ph/0406335}}, \href
  {http://dx.doi.org/10.1016/j.nuclphysbps.2004.09.041}
  {\path{doi:10.1016/j.nuclphysbps.2004.09.041}}.

\bibitem{Hahn:2000kx}
T.~Hahn, {Generating Feynman diagrams and amplitudes with FeynArts 3}, Comput.
  Phys. Commun. 140 (2001) 418--431.
\newblock \href {http://arxiv.org/abs/hep-ph/0012260}
  {\path{arXiv:hep-ph/0012260}}, \href
  {http://dx.doi.org/10.1016/S0010-4655(01)00290-9}
  {\path{doi:10.1016/S0010-4655(01)00290-9}}.

\bibitem{Smirnov:2019qkx}
A.~V. Smirnov, F.~S. Chuharev, {FIRE6: Feynman Integral REduction with Modular
  Arithmetic}, Comput. Phys. Commun. 247  (2020) 106877.
\newblock \href {http://arxiv.org/abs/1901.07808} {\path{arXiv:1901.07808}},
  \href {http://dx.doi.org/10.1016/j.cpc.2019.106877}
  {\path{doi:10.1016/j.cpc.2019.106877}}.

\bibitem{Lee:2013mka}
R.~N. Lee, {LiteRed 1.4: a powerful tool for reduction of multiloop integrals},
  J. Phys. Conf. Ser. 523 (2014) 012059.
\newblock \href {http://arxiv.org/abs/1310.1145} {\path{arXiv:1310.1145}},
  \href {http://dx.doi.org/10.1088/1742-6596/523/1/012059}
  {\path{doi:10.1088/1742-6596/523/1/012059}}.

\bibitem{DiVita:2017xlr}
S.~Di~Vita, P.~Mastrolia, A.~Primo, U.~Schubert, {Two-loop master integrals for
  the leading QCD corrections to the Higgs coupling to a $W$ pair and to the
  triple gauge couplings $ZWW$ and $\gamma^*WW$}, JHEP 04 (2017) 008.
\newblock \href {http://arxiv.org/abs/1702.07331} {\path{arXiv:1702.07331}},
  \href {http://dx.doi.org/10.1007/JHEP04(2017)008}
  {\path{doi:10.1007/JHEP04(2017)008}}.

\bibitem{Ma:2021cxg}
C.~Ma, Y.~Wang, X.~Xu, L.~L. Yang, B.~Zhou, {Mixed QCD-EW corrections for Higgs
  leptonic decay via $HW^{+}W^{-}$ vertex}, JHEP 09 (2021) 114.
\newblock \href {http://arxiv.org/abs/2105.06316} {\path{arXiv:2105.06316}},
  \href {http://dx.doi.org/10.1007/JHEP09(2021)114}
  {\path{doi:10.1007/JHEP09(2021)114}}.

\bibitem{ParticleDataGroup:2020ssz}
P.~A. Zyla, et~al., {Review of Particle Physics}, PTEP 2020~(8) (2020) 083C01.
\newblock \href {http://dx.doi.org/10.1093/ptep/ptaa104}
  {\path{doi:10.1093/ptep/ptaa104}}.

\bibitem{Hahn:2004fe}
T.~Hahn, {CUBA: A Library for multidimensional numerical integration}, Comput.
  Phys. Commun. 168 (2005) 78--95.
\newblock \href {http://arxiv.org/abs/hep-ph/0404043}
  {\path{arXiv:hep-ph/0404043}}, \href
  {http://dx.doi.org/10.1016/j.cpc.2005.01.010}
  {\path{doi:10.1016/j.cpc.2005.01.010}}.

\bibitem{Hahn:2016ktb}
T.~Hahn, {Concurrent Cuba}, Comput. Phys. Commun. 207 (2016) 341--349.
\newblock \href {http://dx.doi.org/10.1016/j.cpc.2016.05.012}
  {\path{doi:10.1016/j.cpc.2016.05.012}}.

\bibitem{Shtabovenko:2016olh}
V.~Shtabovenko, {FeynCalc 9}, J. Phys. Conf. Ser. 762~(1) (2016) 012064.
\newblock \href {http://arxiv.org/abs/1604.06709} {\path{arXiv:1604.06709}},
  \href {http://dx.doi.org/10.1088/1742-6596/762/1/012064}
  {\path{doi:10.1088/1742-6596/762/1/012064}}.

\bibitem{Kallweit:2014xda}
S.~Kallweit, J.~M. Lindert, P.~Maierh\"ofer, S.~Pozzorini, M.~Sch\"onherr, {NLO
  electroweak automation and precise predictions for W+multijet production at
  the LHC}, JHEP 04 (2015) 012.
\newblock \href {http://arxiv.org/abs/1412.5157} {\path{arXiv:1412.5157}},
  \href {http://dx.doi.org/10.1007/JHEP04(2015)012}
  {\path{doi:10.1007/JHEP04(2015)012}}.

\bibitem{Sherpa:2019gpd}
E.~Bothmann, et~al., {Event Generation with Sherpa 2.2}, SciPost Phys. 7~(3)
  (2019) 034.
\newblock \href {http://arxiv.org/abs/1905.09127} {\path{arXiv:1905.09127}},
  \href {http://dx.doi.org/10.21468/SciPostPhys.7.3.034}
  {\path{doi:10.21468/SciPostPhys.7.3.034}}.

\bibitem{Buccioni:2019sur}
F.~Buccioni, J.-N. Lang, J.~M. Lindert, P.~Maierh\"ofer, S.~Pozzorini,
  H.~Zhang, M.~F. Zoller, {OpenLoops 2}, Eur. Phys. J. C 79~(10) (2019) 866.
\newblock \href {http://arxiv.org/abs/1907.13071} {\path{arXiv:1907.13071}},
  \href {http://dx.doi.org/10.1140/epjc/s10052-019-7306-2}
  {\path{doi:10.1140/epjc/s10052-019-7306-2}}.

\bibitem{Adams:2016xah}
L.~Adams, C.~Bogner, A.~Schweitzer, S.~Weinzierl, {The kite integral to all
  orders in terms of elliptic polylogarithms}, J. Math. Phys. 57~(12) (2016)
  122302.
\newblock \href {http://arxiv.org/abs/1607.01571} {\path{arXiv:1607.01571}},
  \href {http://dx.doi.org/10.1063/1.4969060} {\path{doi:10.1063/1.4969060}}.

\bibitem{Remiddi:2017har}
E.~Remiddi, L.~Tancredi, {An Elliptic Generalization of Multiple
  Polylogarithms}, Nucl. Phys. B 925 (2017) 212--251.
\newblock \href {http://arxiv.org/abs/1709.03622} {\path{arXiv:1709.03622}},
  \href {http://dx.doi.org/10.1016/j.nuclphysb.2017.10.007}
  {\path{doi:10.1016/j.nuclphysb.2017.10.007}}.

\bibitem{Broedel:2017kkb}
J.~Broedel, C.~Duhr, F.~Dulat, L.~Tancredi, {Elliptic polylogarithms and
  iterated integrals on elliptic curves. Part I: general formalism}, JHEP 05
  (2018) 093.
\newblock \href {http://arxiv.org/abs/1712.07089} {\path{arXiv:1712.07089}},
  \href {http://dx.doi.org/10.1007/JHEP05(2018)093}
  {\path{doi:10.1007/JHEP05(2018)093}}.

\bibitem{Broedel:2018qkq}
J.~Broedel, C.~Duhr, F.~Dulat, B.~Penante, L.~Tancredi, {Elliptic Feynman
  integrals and pure functions}, JHEP 01 (2019) 023.
\newblock \href {http://arxiv.org/abs/1809.10698} {\path{arXiv:1809.10698}},
  \href {http://dx.doi.org/10.1007/JHEP01(2019)023}
  {\path{doi:10.1007/JHEP01(2019)023}}.

\end{thebibliography}


\begin{thebibliography}{0}
\bibitem{Goncharov}
A.~B. Goncharov, \emph{{Multiple polylogarithms, cyclotomy and modular
  complexes}}, \href{https://doi.org/10.4310/MRL.1998.v5.n4.a7}{\emph{Math.
  Res. Lett.} {\bfseries 5} (1998) 497}
  [\href{https://arxiv.org/abs/1105.2076}{{\ttfamily 1105.2076}}].

\bibitem{Vollinga}
J.~Vollinga and S.~Weinzierl, \emph{{Numerical evaluation of multiple
  polylogarithms}},
  \href{https://doi.org/10.1016/j.cpc.2004.12.009}{\emph{Comput. Phys. Commun.}
  {\bfseries 167} (2005) 177}
  [\href{https://arxiv.org/abs/hep-ph/0410259}{{\ttfamily hep-ph/0410259}}].                           
\end{thebibliography}







\end{document}